\documentclass[preprint,showpacs,showkeys,aps,prb]{revtex4}
\usepackage[english]{babel}

\usepackage[dvips]{graphicx}
\usepackage{amssymb}

\begin{document}


\title{Nonlinear Stresses and Temperatures in Transient Adiabatic and
Shear Flows {\em via} Nonequilibrium Molecular
Dynamics -- Three Definitions of Temperature.
}

\author{Wm. G. Hoover and C. G. Hoover \\
Ruby Valley Research Institute \\ Highway Contract 60,
Box 598, Ruby Valley 89833, NV USA    }
\date{\today}

\pacs{02.70.Ns, 45.10.-b, 46.15.-x, 47.11.Mn, 83.10.Ff}


\keywords{Thermostats, Ergostats, Molecular Dynamics, Computational Methods,
Smooth Particles}

\vskip 0.5cm

\begin{abstract}

We compare nonlinear stresses and temperatures for adiabatic shear flows,
using up to 262,144 particles, with those from corresponding homogeneous
and inhomogeneous flows.  Two varieties of kinetic temperature tensors
are compared to the configurational temperatures. This comparison  of
temperatures led us to two new findings, beyond our original goal of
analyzing shear algorithms.  First, we found an improved form for local
instantaneous velocity fluctuations, as calculated with smooth-particle
weighting functions.  Second, we came upon the previously unrecognized
contribution of rotation to the configurational temperature.

\end{abstract}

\maketitle

\section{Introduction}

Hoover, Hoover, and Petravic\cite{b1} studied the nonlinear stresses and
temperature changes induced by shear in a variety of stationary flows.
They used nonequilibrium molecular dynamics to compare several popular
algorithms.  The algorithms are described briefly in Sec. II.  See
Figs. 1 and 2 for the geometries used to induce the flows.  This work used
a smooth repulsive soft-sphere potential\cite{b2} with a range of unity,
$$
\phi(r<1) = 100(1-r^2)^4 \ .
$$
The particle mass, energy per particle, and density were all chosen equal
to unity.  These conditions correspond to a dense fluid at about
2/3 the freezing pressure.
$$
m = 1 \ ; \ E/N = (K + \Phi)/N = 1 \ ; \ \rho = Nm/V = 1 \ .
$$
Thermostat or ergostat forces were used to
generate stationary states.  There, as well as in the present work,
we choose $x$ for the flow direction and $\dot \epsilon = 0.5$ for
the strainrate, where the
time-averaged velocity component of the flow $v_x$
increases linearly in the $y$ direction:
$$
\langle v_x(y) \rangle = \dot \epsilon y \ .
$$
Three-dimensional homogeneous periodic simulations [Fig. 1 shows a
two-dimensional version] gave
$$
T_{yy} > T_{xx} > T_{zz} \ ; \
P_{yy} > P_{xx} > P_{zz} \ {\rm [Doll's \ Algorithm]} \
 ;
$$
$$
T_{xx} > T_{yy} > T_{zz} \ ; \
P_{xx} > P_{yy} > P_{zz} \ {\rm [Sllod \ Algorithm]} \ .
$$
and differed qualitatively from corresponding three-dimensional
boundary-driven results [Fig. 2 shows a two-dimensional version]:
$$
T_{xx} > T_{zz} > T_{yy} \ ; \ P_{xx} > P_{zz} > P_{yy}
\ {\rm [Boundary-Driven]} \ .
$$

The main conclusion drawn from that work was that {\em neither} homogeneous
method, Doll's\cite{b3} nor Sllod\cite{b4}, successfully reproduces the
more-physical boundary-driven results\cite{b5}.  To quote Ref. 1: ``The
Doll's and Sllod algorithms predict opposite signs for this
normal-stress difference $[P_{xx}-P_{yy}]$, with the Sllod approach
definitely wrong, but somewhat closer to the (boundary-driven) truth.''

\begin{figure}
\includegraphics[height=6cm,width=6cm,angle=-90]{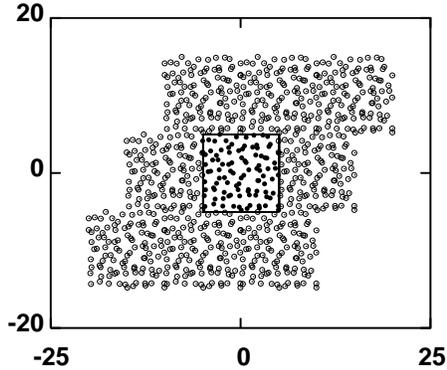}
\caption{
Two-dimensional version of periodic homogeneous isoenergetic shear flow.
Eight periodic images of the central $N$-particle system are shown.
}
\end{figure}

\begin{figure}
\includegraphics[height=8cm,width=6cm,angle=-0]{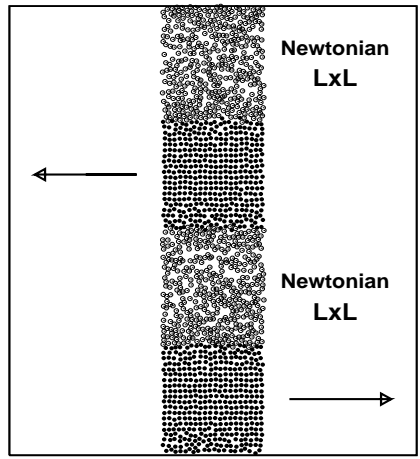}
\caption{
Two-dimensional version of periodic inhomogeneous boundary-driven shear
flow.  The system consists of four separate chambers of $N$ particles each.
The chambers indicated by arrows are driven to the right and left by moving
tether forces.  Heat is extracted from the driven chambers to maintain
constant internal energy there.  The boundary-driven motion of the other
two chambers is purely Newtonian.
}
\end{figure}

Evans objected to this conclusion\cite{b6}, stating that the Sllod algorithm
is ``exact''.  He is of course correct, in the sense that the Sllod algorithm
is nothing more than Newton's equations of motion written in a different
coordinate frame, a Lagrangian frame moving along with the sheared fluid.
But because Newton's equations by themselves cannot lead
to nonequilibrium steady states, the Sllod algorithm is ``exact'' in the
rather limited case of isolated systems.

The misunderstanding evident in
Evans' remark led us to undertake the present work.  Instead of
considering steady states, which seem to us the simplest situation, Evans
had in mind a time-dependent spatially-periodic adiabatic deformation.  When
no thermostat or ergostat forces are used in the equations of motion the
shear deformation is adiabatic, with continuous heating.  In the adiabatic
case no steady state results and the Sllod equations of motion are equivalent
to Newton's equations of motion for a system undergoing periodic deformation
with strainrate
$\dot \epsilon$.  Just as in the thermostated case the normal stresses
and temperatures differ.  Not only the magnitudes, but also the orderings,
of these components can, and do, differ from those found in steady states.

In this paper we motivate and describe large-scale adiabatic-shear
simulations and discuss the interpretation of these simulations.  These
simulations use periodic boundary conditions, just as shown in Fig. 1, but are
extended here to three Cartesian space dimensions $\{x,y,z\}$.  We take
into account the important r\^ole of fluctuations in defining local
values of the velocity, and the temperature and stress tensors.  Sec. II
describes the algorithms, and Sec. III the various definitions of temperature 
for nonequilibrium (as well as equilibrium) systems.  Sec. IV outlines the
results of the current simulations.  Sec. V gives the conclusions we have
reached as a result of this work.  Sec. VI suggest extensions of this work.

\section{Shear Flow Algorithms}

Two numerical algorithms, ``Sllod'' and ``Doll's'', for spatially-periodic
shear flow in a volume $V$ both satisfy the macroscopic energy-balance
relation
$\dot E = -\dot \epsilon P_{xy}V$. The corresponding solutions differ
in effects of order $\dot \epsilon ^2$, with $P_{xx} > P_{yy}$ in the Sllod
case, and the reverse using Doll's algorithm.
 To
describe nonequilibrium situations it is natural to introduce the
gradients and time derivatives of these variables, with the simplest
situations those ``stationary states'' (necessarily driven by external
forces or heat sources) in which all the partial time derivatives (the
rates of change at a fixed location) vanish.  Steady shear flow can be
simulated with homogeneous sources and sinks of momentum and energy
through the Doll's and Sllod algorithms.  The {\em adiabatic} versions
of these equations of motion (no thermostats or ergostats) introduce
an overall flowfield imposed with the parameter $\dot \epsilon$
through periodic boundary conditions:
$$
\dot x \equiv (p_x/m) + \dot \epsilon y \ ; \ \dot y = (p_y/m) \ ; \ 
 {\rm [Sllod \ or \ Doll's]} \ .
$$
The Sllod algorithm is simply a rewriting of Newton's equations of motion
for the evolution of the ``momentum'' $p_x$ relative to the motion induced
by the periodic boundary conditions:
$$
\ddot x = (F_x/m) \ ; \ \ddot y =  (F_y/m) \
\longleftrightarrow
$$
$$
m \ddot x - m\dot \epsilon \dot y = \dot p_x = F_x - \dot \epsilon p_y  \ ; \
m \ddot y = \dot p_y = F_y   \ {\rm [Sllod]} \ .
$$

In the laboratory frame (where one sees the overall strainrate $\dot \epsilon$
induced by the periodic boundary conditions) the motion follows from the usual
Hamiltonian,
$$
{\cal H}_{\rm Lab} = \sum p^2/(2m) + \Phi \ .
$$

If, as it is in the Sllod algorithm, the momentum $(p_x,p_y)$ is
defined instead in the comoving frame then there is no analogous Hamiltonian.
To see this in detail suppose that the comoving equations of motion
(describing the Newtonian dynamics) could be derived from a hypothetical
comoving Hamiltonian, ${\cal H}_{\rm com}(\{ x,y,p_x,p_y\} )$:
$$
\dot x = +(\partial {\cal H}_{\rm com}/\partial p_x)
= (p_x/m) + \dot \epsilon y \ ; \
\dot y = +(\partial {\cal H}_{\rm com}/\partial p_y) = (p_y/m) \ ;
$$
$$
\dot p_x = -(\partial {\cal H}_{\rm com}/\partial x)
= F_x - \dot \epsilon p_y  \ ; \
\dot p_y = -(\partial {\cal H}_{\rm com}/\partial y) = F_y  \ .
$$
The second partial derivatives of the hypothetical Hamiltonian with respect
to $y$ and $p_x$ should be equal.  But we find instead
$$
(\partial /\partial y)(\partial {\cal H}_{\rm com}/\partial p_x) =
(\partial /\partial y)(\dot \epsilon y) = \dot \epsilon \ ,
$$
and
$$
(\partial /\partial p_x)(\partial {\cal H}_{\rm com}/\partial y) =
(\partial /\partial p_x)(-F_y) = 0 \ ,
$$
showing that there is no such comoving Hamiltonian.

On the other hand, the very similar Doll's-Tensor equations of motion
(which are not Newtonian)
do follow from a special Hamiltonian appropriate to the comoving frame:
$$
{\cal H}_{\rm Doll's} = \sum p^2/(2m) + \Phi + \dot \epsilon \sum yp_x \ :
$$
$$
+(\partial {\cal H}_{\rm Doll's}/\partial p_x) = \dot x =
(p_x/m) + \dot \epsilon y  \ ; \
+(\partial {\cal H}_{\rm Doll's}/\partial p_y) = \dot y =
(p_y/m) \ .
$$
$$
-(\partial {\cal H}_{\rm Doll's}/\partial x) = \dot p_x = F_x  \ ; \
-(\partial {\cal H}_{\rm Doll's}/\partial y) = \dot p_y = F_y - \dot \epsilon
p_x \ .
$$

Both the foregoing Sllod and Doll's sets of motion equations are adiabatic,
so that the systems they describe heat due to viscous shear as time goes on.
Additional time-reversible frictional forces of the form $-\zeta p$ can
be added to either set of motion equations to keep the energy or the
temperature constant\cite{b7,b8}:
$$
\{ \ \Delta F = - \zeta p \ ; \ p_x \equiv m(\dot x - \dot \epsilon y) \ ; \
p_y = m\dot y \ ; \ p_z = m\dot z \ \} \ .
$$
The frictional forces make it possible to explore a spatially-homogeneous
nonequilibrium steady state with definite values of the (time-averaged)
stress and temperature.  Both these nonequilibrium properties need proper
definitions.  We consider several alternative definitions of temperature
in the following section. 

\section{Definitions of Temperature}

In statistical mechanics a longstanding definition of temperature has been
kinetic, based on the physical picture of an ideal-gas
thermometer\cite{b7,b9}.
Measuring the momenta $\{ p \}$ relative to the comoving frame of the
kinetic thermometer, the kinetic-theory definition is:
$$
kT_{xx} \equiv \langle p_x^2/m \rangle \ ; \
kT_{yy} \equiv \langle p_y^2/m \rangle \ ; \
kT_{zz} \equiv \langle p_z^2/m \rangle \ .
$$
A simple mechanical model capable of measuring all three temperatures
simultaneously is a dilute gas of parallel hard cubes\cite{b10}.

There is also a configurational analog\cite{b11,b12},
$$
kT_{xx} = \langle F^2_x \rangle / \langle \nabla _x^2{\cal H} \rangle \ ; \
kT_{yy} = \langle F^2_y \rangle / \langle \nabla _y^2{\cal H} \rangle \ ; \
kT_{zz} = \langle F^2_z \rangle / \langle \nabla _z^2{\cal H} \rangle \ .
$$
The configurational temperature has no clear connection to a physical
model of a thermometer, but follows instead\cite{b11} from a formal
integration by parts of the canonical average of $\nabla ^2{\cal H}$.
$$
\langle \nabla ^2 {\cal H} \rangle \equiv 
\langle (\nabla {\cal H})^2/kT \rangle \  .
$$

Both temperature definitions, kinetic and configurational, have
associated ambiguities: [1] fluctuations in the kinetic case,
and [2] rotation in the configurational case.  Consider
fluctuations first.  The local velocity fluctuates in time.  The thermal
momentum in the kinetic definition
has to be measured in a ``comoving'' frame.  Once the velocity is a
local quantity, as well as a time-dependent quantity, its definition
becomes crucial.  Here we adopt a modification of the ``smooth-particle''
definition of local velocity\cite{b13}:
$$
v(r) \equiv \sum w(|r-r_i|<h)v_i / \sum w(|r-r_i|<h) \ ,
$$
where $v_i$ is the velocity of Particle i, and that particle lies within
the range $h$ of the smooth-particle weighting function $w(r<h)$.

SPAM [Smooth Particle Applied Mechanics\cite{b13}] provides spatially
very smooth material properties (such as density, velocity, stress, and
energy) with two continuous spatial derivatives.  The definitions of these
properties require a smooth weighting function, $w(r<h)$, which must be
continuously twice differentiable, normalized, and which must also have
a finite range $h$.  Here we adopt the simplest such weighting function
meeting these requirements, Lucy's.  In three dimensions Lucy's form for
$w$ is
$$
w_{\rm Lucy} = \frac{105}{16\pi h^3}[1 - 6x^2 + 8x^3 - 3x^4] \ ; \
x \equiv |r|/h \ ;
$$
$$
 \rightarrow \int_0^\infty 4\pi r ^2w(r<h)dr \equiv
 \int_0^h 4\pi r ^2w(r<h)dr \equiv 1 \ .
$$
In the following section we show that the smooth particulate velocity
fluctuations measured as temperature are best defined through a slight
modification of the smooth-particle approach, in which the ``self''
contributions to the particle sums, $\sum w$ and $\sum wv$, are absent.
This modification reduces the number-dependence inherent in comparing
atomistic simulations to continuum predictions.

At first sight, the configurational definition of temperature has an
advantage over the kinetic one in that a calculation of the stream
velocity is not required.  But the current work led us to recognize
a difficulty in defining configurational temperature away from
equilibrium.  Consider rotation.  Particularly in turbulent flows,
{\em rotation} is important.  Although
configurational temperature has been touted as a way to avoid
defining a local velocity\cite{b13}, it also contains a
small and subtle ambiguity --- configurational temperature depends on
rotation rate.

A rotating rigid body generates centrifugal forces of
order $\omega ^2r$ (offset by tensile forces) at a distance $r$ from
the center of mass. The tensile forces contribute to the configurational
temperature definition,
$$
kT_C \equiv \langle F^2\rangle / \langle \nabla ^2{\cal H} \rangle \ ,
$$
while the centrifugal ones do not, so that perimeter particles are
apparently ``hotter'' than the cooler interior by (relatively-small)
contributions of order $\omega ^4$.

To see this in a simple two-dimensional example, consider the point
$(x,y)$ viewed from an $(X,Y)$ coordinate system rotating counterclockwise
at the angular frequency $\omega $:
$$
X = x\cos (\omega t) + y\sin (\omega t) \ ; \
Y = y\cos (\omega t) - x\sin (\omega t) \ .
$$
Two time differentiations, evaluated at time $t=0$, give the Coriolis,
centrifugal (rotating frame), and centripetal (laboratory frame)
forces:
$$
\ddot X = \ddot x + 2\omega \dot y - \omega ^2 x =
          \ddot x + 2\omega \dot Y + \omega ^2 X \ ; 
$$
$$
\ddot Y = \ddot y - 2\omega \dot x - \omega ^2 y =
          \ddot y - 2\omega \dot X + \omega ^2 Y \ .
$$
For rigid rotation at an angular velocity $\omega$ a particle at
$(x,y) = (r,0)$ with laboratory frame velocity $(0,\omega r)$ and
acceleration $(F_x,0)/m = (-\omega ^2 r,0)$ the $\ddot X$ equation
becomes:
$$
\ddot X = \ddot x + 2\omega \dot y - \omega ^2 x \ =
(F_x/m) + 2\omega ^2 r - \omega ^2 r \equiv 0 = (F_x/m) + \omega ^2 r \ ,
$$
showing that the atomistic force $(F_x/m)$ exactly offsets the
centrifugal force $\omega ^2 r$.  Thus the configurational temperature
for rigid rotation is proportional to $r^2\omega ^4$.

The numerical work described in the following section, for relatively
gentle shear flows, supports the view that kinetic temperature is both
simpler and better behaved than configurational temperature, with smaller
fluctuations in both space and time.  In a work still in progress we
contrast the two approaches for the problem of a strong shockwave.

\section{Simulations and Results}

Throughout, we focus on a dense fluid in three space dimensions, with the
short-ranged ``soft-sphere'' potential of Refs. 1 and 2 :
$$
\phi(r<1) = 100(1 - r^2)^4 \ ; \ \Phi = \sum _{i<j} \phi(|r_{ij}|) \ ,
$$
where the sum over pairs includes all particle pairs within a
distance unity.  The total energy of the system consists of a
kinetic part in addition to the potential energy $\Phi $.
$$
E = K + \Phi \ ; \ K = \sum p^2/(2m) \ .
$$
We focus on the dense-fluid state of Ref. 1, with a density and energy per
particle of unity,
$$
E/N = Nm/V = N/V = 1 \ ; \ 6^3 = 216 \leq N \leq 2744 = 14^3 \ .
$$

\begin{figure}
\includegraphics[height=10cm,width=6cm,angle=-90]{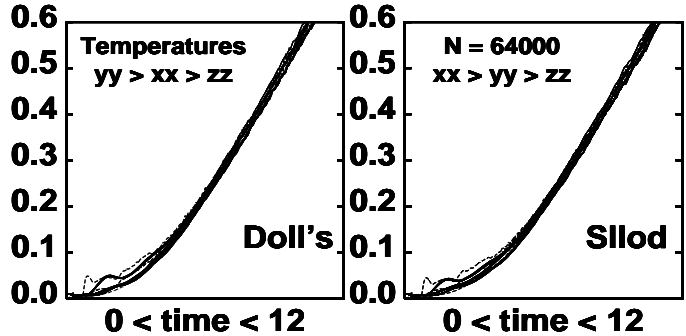}
\caption{
Overall adiabatic temperature variations for adiabatic shear flows with
$64 \times 64 \times 64$ soft spheres with an initial kinetic temperature
of 0.01.  The strainrate $du_x/dy = \dot \epsilon $ is 0.5.
$\{T_{xx},T_{yy},T_{zz}\}$  are plotted here.  For times greater than 3
neither algorithm shows significant differences between the temperatures
on the scale of the plots.
}
\end{figure}

\begin{figure}
\includegraphics[height=10cm,width=6cm,angle=-90]{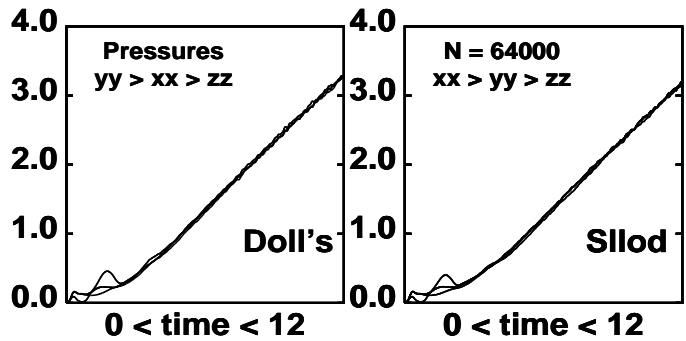}
\caption{
Overall adiabatic pressure variation for adiabatic shear flows with
$64 \times 64 \times 64$ soft spheres with an initial kinetic temperature
of 0.01.  The strainrate $du_x/dy = \dot \epsilon $ is 0.5.
$\{P_{xx},P_{yy},P_{zz}\} $ are plotted here.
For times greater than 3 neither algorithm shows significant differences
between the pressures on the scale of the plots.
}
\end{figure}

Data for homogeneous isoenergetic Doll's and Sllod simulations are
given in Ref. 1 along with complementary results for boundary-thermostated
flows.  That study
showed that the normal stress differences in the homogeneous
simulations are very different to those found in boundary-driven flows.
The number-dependence in the temperatures and normal stress differences
of the homogeneous flows is nearly negligible, no more than $1/N$ once
the number of particles $N$ is a few hundred.  By contrast, the
boundary-driven temperatures are quite different, as the midstream
temperature increases as $N^{2/3}$.

Here we consider in addition {\em adiabatic deformation}, with the
Newtonian motion driven by shearing boundary conditions and without
any thermostat or ergostat forces.  The initial state is a cubic
lattice with a kinetic temperature of $kT = 0.01$ and an initial energy per
particle of $E=K + \Phi = 0.015N + 0$ (because the nearest-neighbor
separation is initially unity, just beyond the range of the repulsive
forces).  We compute and compare two different kinetic
temperatures, each with the three components
$\{ T_{xx},T_{yy},T_{zz} \}$.  The {\em time-averaged temperature},
$kT^{\rm TA}$ is
$$
kT_{xx}^{\rm TA} \equiv \langle m(\dot x_i - \dot \epsilon y_i)^2 \rangle \
;
$$
$$
kT_{yy}^{\rm TA} \equiv \langle m\dot y^2 \rangle \ ;
$$
$$
kT_{zz}^{\rm TA} \equiv \langle m\dot z^2 \rangle \ ,
$$
while the instantantaneous temperature, $kT^{\rm inst}$ is
$$
kT_{xx}^{\rm inst} \equiv \langle m(\dot x_i - v_x(r_i,t))^2 \rangle \ ; \
$$
$$
kT_{yy}^{\rm inst} \equiv \langle m(\dot y_i - v_y(r_i,t))^2 \rangle \ ; \
$$
$$
kT_{zz}^{\rm inst} \equiv \langle m(\dot z_i - v_z(r_i,t))^2 \rangle \ ,
$$
where the instantaneous velocity at Particle i's location is a modified
version of the usual smooth-particle average\cite{b13}:
$$
v(r_i,t) = \sum_{j\neq i}w(r_{ij})v_j/ \sum_{j\neq i}w(r_{ij}) \ .
$$
For simplicity we choose Lucy's weight function with the range $h=3$ :
$$
w(r<h) = \frac{105}{16\pi h^3}(1 - 6x^2 + 8x^3 - 3x^4) \ ; \ x \equiv r/h \ ,
$$
for the evaluation of all the smooth-particle sums.

\begin{figure}
\includegraphics[height=10cm,width=6cm,angle=-90]{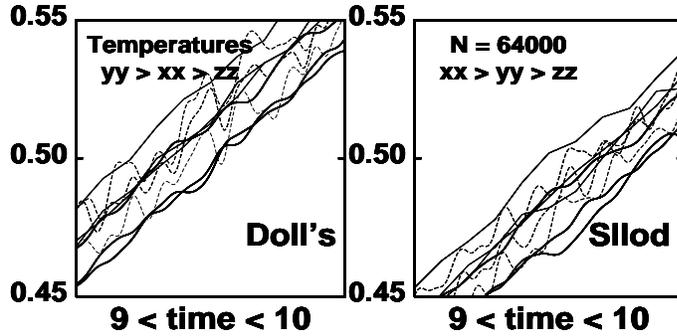}
\caption{
$\{T_{xx},T_{yy},T_{zz}\} $ are plotted here for  portions of 64,000-particle
adiabatic simulations of Figs. 3 and 4. The heaviest lines show the 
laboratory-frame kinetic temperature; the medium lines show kinetic
temperature relative to the instantaneous smooth-particle velocity.  The
light dashed lines show the configurational temperatures, which fluctuate more
wildly than the kinetic temperatures.  In the steady-state simulations
of Ref. 1 the Doll's kinetic temperatures $\{0.496,0.508,0.493\}$ and
the Sllod kinetic temperatures $\{0.507,0.497,0.493\}$ correspond to an
internal per-particle energy of exactly unity.
}
\end{figure}

\begin{figure}
\includegraphics[height=10cm,width=6cm,angle=-90]{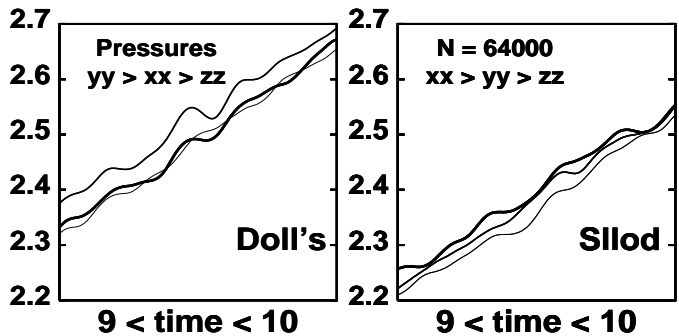}
\caption{
$\{P_{xx},P_{yy},P_{zz}\} $ are plotted here for  portions of the 
adiabatic simulations of Figs. 3 and 4 (using laboratory-frame
kinetic contributions) and correspond to the heavy, medium, and light lines
respectively. In the steady-state simulations of Ref. 1 the Doll's pressures
(using laboratory-frame kinetic parts) $\{2.496,2.528,2.482\}$ and the Sllod
pressures $\{2.516,2.509,2.484\}$ correspond to an internal per-particle
energy of exactly unity and an average temperature of about 0.5.  The shear
stress is about the same for the two algorithms, $\sigma _{xy} = 0.343$.
}
\end{figure}

In smooth-particle simulations\cite{b13} the ``self terms'', in the two
particle sums, $w_{ii}v_i$ and $w_{ii}$ are always included.  In analyzing
equilibrium molecular dynamics simulations for local velocity fluctuations
(the usual kinetic temperature) numerical work shows that there is a much
better correspondence with equilibrium temperature when the self terms
are omitted.  We have followed that practice here.  If the self terms are
included in computing the local stream velocity the resulting kinetic
temperatures are roughly 10$\%$ lower.  With the self terms omitted the
three temperature definitions, time-averaged kinetic, instantaneous
kinetic, and configurational, all give similar results. 

The need for excluding the ``self'' contributions can be rationalized by
considering an equilibrium particle $i$ at location $r_i$ with velocity
$v_i$.  With its neighbors' velocities uncorrelated (as they are at
equilibrium) the smooth-particle velocity at $r_i$ is, on average,
$$
v_{\rm SPAM}(r_i) \equiv \sum _jw_{ij}v_j/\sum _jw_{ij} \simeq w(0)v_i \ .
$$ 
so that the temperature, based on the velocity fluctuations as 
measured by the differences, $\{ v_i - \langle v_{\rm SPAM}(r_i)\rangle \} $,
is reduced by a factor of $[1-w(0)]^2$.  Accordingly, we have excluded
the ``self terms'' in the kinetic parts of the temperatures and pressures
illustrated in the figures.

 Figs. 3 and
4 show the overall increase of temperature and pressure beginning with a
homogeneous cubic crystal, at a kinetic temperature of 0.01, and ending
at a homogeneous shearing fluid state with a temperature somewhat greater
than 0.5.  The details of the temperature and pressure for 64,000 particles,
in the vicinity of $kT \simeq 0.5$, are shown in Figs. 5 and 6.  The
fluctuations in the data can be reduced by using even larger systems.  Compare
Figs. 5 and 6 with corresponding results for 262,144 particles, shown in
Figs. 7 and 8.  In these latter simulations the internal energy per particle
reaches unity for the Sllod algorithm at a time of 9.799, and for the Doll's
algorithm at a time of 9.480.

At a fixed strainrate of 0.5, small-system fluctuations can completely obscure
the orderings of $\{ T_{ii} \} $ and $\{ P_{ii} \} $.  By increasing the system
size it is possible to verify that the transient fluctuating temperatures
and stresses in adiabatic deformation are close to those of the isoenergetic
periodic shears, with the orderings $y>x>z$ for Doll's and $x>y>z$ for
Sllod.  Neither algorithm reproduces the boundary-driven ordering (at the
same density, strainrate, and energy) $x>z>y$.  We discuss this finding 
in the following Section.

\begin{figure}
\includegraphics[height=10cm,width=6cm,angle=-90]{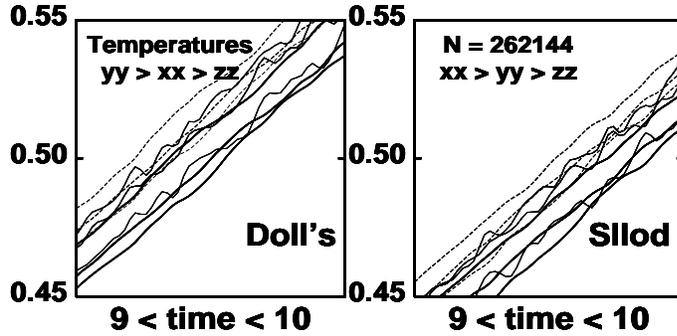}
\caption{
$\{T_{xx},T_{yy},T_{zz}\} $ are plotted here for portions of 262,144-particle
adiabatic shear simulations. The heaviest lines show the 
laboratory-frame kinetic temperature; the medium lines show kinetic
temperature relative to the instantaneous smooth-particle velocity.  The
light dashed lines show the configurational temperatures, which fluctuate more
than do the kinetic temperatures. The kinetic temperatures are nearly
the same as those in the stationary shear simulations of Ref. 1.
} 
\end{figure}

\begin{figure}
\includegraphics[height=10cm,width=6cm,angle=-90]{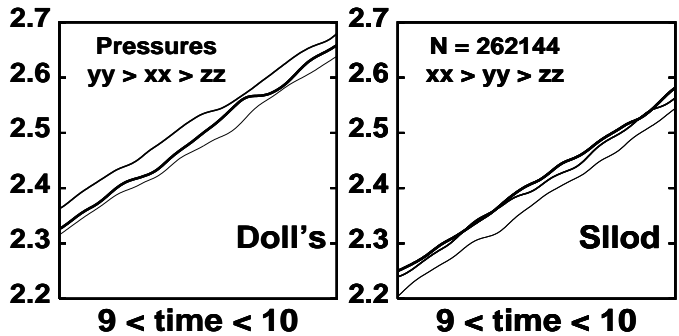}
\caption{
$\{P_{xx},P_{yy},P_{zz}\} $ are plotted here for portions of
262,144-particle adiabatic shear simulations, and correspond to the
heavy, medium, and light lines respectively.  The configurational
parts of the pressure are slightly, but significantly, larger than
those found in the stationary shear simulations of Ref. l.
}
\end{figure}

\section{Nonequilibrium Constitutive Relations}

Models for continuum mechanics follow from conservation of mass,
momentum, and energy.  The differential expressions of these conservation
relations are the continuity equation, the equation of motion
(which introduces the pressure tensor $P$ as the comoving momentum flux),
and the energy equation (which introduces the heat flux vector $Q$ as the
comoving energy flux):
$$
\dot \rho = - \rho \nabla \cdot v \ ; \
\rho \dot v = - \nabla \cdot P \ ; \
\rho \dot e = - \nabla v : P - \nabla \cdot Q \ .
$$
The time-and-space dependent state variables of hydrodynamics
are taken from equilibrium thermodynamics, extended to the case in which
gradients and time dependence can occur.  The state variables at the
location $r$ and time $t$ are the density, velocity, and energy,
$\{ \rho (r,t), v(r,t), e(r,t) \}$ and it is assumed that the pressure
and heat flux can be defined in terms of the present values, the
gradients, and possibly the past histories, of these same variables. 

In the present work we have seen that both the temperature (extended from
the scalar thermodynamic variable to tensor values) and the stress can
differ for two systems with identical densities, strainrates, energies,
and constitutive relations (because the underlying particles are the
same).  Evidently both temperature and stress depend upon additional
state variables.  The relative independence of the normal stress
differences to the system size\cite{b1} $L$ suggests that the discrepancy
between periodic and boundary-driven systems is insensitive to second
derivatives,
$\{\nabla \nabla \rho, \nabla \nabla v,\nabla \nabla e\}$, all of
which vary as $L^{-2}$.  From the constitutive standpoint it is simplest
to imagine a dependence of the normal stress differences and the
temperature tensor on the rate of heating $\dot e$. Such a dependence
could be used to describe the deviation of the
adiabatic transient flows from corresponding stationary flows.  Finding
an additional independent constitutive variable to
distinguish stationary boundary-driven flows from stationary
homogeneous flows is a challenging research goal.

\section{Summary}

Evans' emphasis on the exactness of the Sllod algorithm (restricted to
adiabatic flows) is confirmed here, as Sllod is nothing but Newton in
a different coordinate frame.  But it must be noted that the
large-system adiabatic pressure tensor exhibits clear differences
from the stationary pressure tensor at the same energy, density, and
strain rate. Sufficiently large systems, with hundreds of
thousands of particles, show that the nonequilibrium temperature
tensors of adiabatic transient flows are very similar to
those of homogeneous periodic stationary flows.  The ``realism'' of 
the adiabatic flows is questionable because real boundaries,
which normally drive, constrain, and cool flows, are absent.  The
diffusion time
for an $N$-particle system driven by a strainrate incorporated in its
periodic boundary conditions varies as $N^{2/3}$, so that simulation
results depend increasingly upon their initial conditions as system
size increases.

An interesting finding of the present work is that the smooth-particle
calculation of local velocity (needed for the computation of the local
temperature), $v_{\rm SPAM}=\sum wv/\sum w$, is best modified by
omitting the self terms in both sums.  Considering the fluctuations in
$v - \langle v_{\rm SPAM}\rangle $ in an equilibrium system, we see that
omitting the self terms results in an {\em exactly correct} temperature,
while including them leads to errors of the order of ten percent.

From the constitutive standpoint it is simplest to ``explain'' the
difference between the adiabatic and boundary-driven nonlinear properties
through a dependence on $\dot e$, where $e$ is the internal energy per
unit mass.  Though the configurational temperature tensor avoids the
problem of defining a local stream velocity it still includes the effect
of rotational contributions, giving rise to ``temperature gradients''
based on centrifugal forces in the absence of heat flow.  Fortunately these
rotational temperature contributions are small, of order $\dot \epsilon ^4$.

\section{What to do?}

One referee asked us Lenin's famous question with regard to this work's
consequences.  We hope to stimulate further investigations of microscopic
systems from macroscopic points of view.  The microscopic analogs of
macroscopic temperature, stress, and fluctuations are imperfect, but
vital in drawing macroscopic conclusions from particle simulations.
There is much to do in understanding this correspondence better for more
complex systems with rotational and vibrational degrees of freedom.
Simulations and theories of elongational flow have led to unresolved
controversies as to the ``right way'' to simulate such flows.  See,
for instance Refs. 28-36 cited in our Ref. 1.  We believe that the
nonlinear aspects of steady deformational flows deserve more study.
For unsteady flows even an ``exact'' algorithm such as Sllod, depends in an
essential way on the initial conditions unless the deformation rate is
very small. 
 
Shockwaves provide more extreme tests of the correspondence between
microscopic and macroscopic models.  The significance of temperature for
quantum systems away from equilibrium needs elucidation too. We are
confident that progress along all of these lines can best be achieved by
carrying out, analyzing, and comparing series of simulations such as
those described in the present work.

\newpage

\section{Acknowledgments}

We thank Denis Evans for his stimulating comments and Carl Dettmann for
his useful remarks on the lack of a Sllod Hamiltonian.  We are grateful
to the two anonymous referees for their constructive criticism.

\end{document}